\documentclass[a4paper,11pt]{article}
\usepackage{pos}
\usepackage{float}
\title{Identifying light charged Higgs boson in the $\mu\nu$ channel in 2HDM Type-III}

\author[a]{Rachid Benbrik}
\author*[a]{Mohammed Boukidi}

\affiliation[a]{Polydisciplinary Faculty, Laboratory of Fundamental and Applied Physics,\\ Cadi Ayyad University, Sidi Bouzid, B.P. 4162, Safi, Morocco}

\emailAdd{r.benbrik@uca.ac.ma}
\emailAdd{mohammed.boukidi@ced.uca.ma}

\abstract{In this contribution, we discuss the light charged Higgs boson production via $pp \to \bar{t} bH^\pm$ at the Large Hadron Collider (LHC) in the Two-Higgs Doublet Model (2HDM) Type-III.  We explore the prospect of looking the aforementioned Higgs boson production channel followed by $H^\pm\mu\nu$ signal. The latter has the potential to be overwhelmingly stronger than  $H^\pm\tau\nu$ and $H^\pm c\bar{s}$ signals in Type-III. We show that in both scenarios standard and inverted hierarchy and after including several theoretical and experimental constraints, the production process $pp \to \bar{t}bH^\pm$ followed by $H^\pm\to \mu\nu$ could represent the most promising experimental option to search for light charged Higgs boson at the LHC.}

\FullConference{%
  The Tenth Annual Conference on Large Hadron Collider Physics - LHCP2022\\
  16-20 May 2022\\
  online
}

\begin{document}
\maketitle
\section{Introduction}
The discovery of a scalar particle at the Large Hadron Collider (LHC) in 2012 \cite{Aad:2012tfa,Chatrchyan:2012ufa} confirmed the Standard Model (SM) as the fundamental theory of strong and electroweak interactions.
However, despite its high compatibility with experimental measurements, the SM has significant shortcomings, and there are several phenomenological indications that the SM cannot answer all questions, so any extension of the SM is well motivated. 

The Two-Higgs-Doublet Model (2HDM) is considered as one of the simplest beyond the SM extension. Apparently the most general implementation of 2HDM has non-diagonal fermionic couplings in the flavour space, and can therefore induce flavor-changing neutral current (FCNC)
effects at the tree level, which might be not consistent with the actual observation. If we allow these tree-level FCNCs, both doublets can couple to leptons and quarks and the associated model is called 2HDM Type-III\footnote{The model is described in detail in  Refs. \cite{Benbrik:2021wyl, Benbrik:2022azi}}, in the latter we assume the so-called Cheng-Sher ansatz \cite{Cheng:1987rs}  within the fermion sector, forcing
the non-diagonal Yukawa couplings to become relative to the mass of the relevant fermions $\tilde{Y}_{ij} \propto \sqrt{m_i m_j}/ v ~\chi_{ij}$, where $\chi_{ij}$  are taken as dimensionless free parameters. Assuming $\mathcal{CP}$-conservation option, and very minimal version of Higgs couplings, the 2HDM can be parameterized by at least 7 free parameters, those are: Higgs masses, $m_h$, $m_H$, $m_{H^\pm}$, $m_A$, the ratio of the vacuum expectation values of the two Higgs doublets fields $\tan\beta=v_2/v_1$, the mixing angle of the CP-even Higgs states $\alpha$, and finally the $m^2_{12}$. 

Charged Higgs boson presents special challenge for experimental search. In this regard ATLAS and CMS collaborations almost target the fermionic decay channels in their programs to look for $H^\pm$. In the $m_{H^\pm}< m_t-m_b$ case the charged Higgs boson has been searching for through the top pair (anti)quark production and decay, $pp \to t \bar{t} \to W^\pm H^\pm b\bar{b}$, with $H^\pm$ decaying
into a pair of fermions $\tau\nu$ and $c\bar{s}$. Our study is dedicated to the $H^\pm \to\mu\nu$ decay to assess the extent to which it might complement the searches for light charged Higgs.

\subsection{Review of 2HDM Type-III}
After spontaneous symmetry breaking the Yukawa Lagrangian can be expressed, in terms of the mass eigenstates of the Higgs bosons, as follows:
\begin{align}
-{\cal L}^{III}_Y  &= \sum_{f=u,d,\ell} \frac{m^f_j }{v} \times\left( (\xi^f_h)_{ij}  \bar f_{Li}  f_{Rj}  h + (\xi^f_H)_{ij} \bar f_{Li}  f_{Rj} H - i (\xi^f_A)_{ij} \bar f_{Li}  f_{Rj} A \right)\nonumber\\  &+ \frac{\sqrt{2}}{v} \sum_{k=1}^3 \bar u_{i} \left[ \left( m^u_i  (\xi^{u*}_A)_{ki}  V_{kj} P_L+ V_{ik}  (\xi^d_A)_{kj}  m^d_j P_R \right) \right] d_{j}  H^+ \nonumber\\  &+ \frac{\sqrt{2}}{v}  \bar \nu_i  (\xi^\ell_A)_{ij} m^\ell_j P_R \ell_j H^+ + H.c.\, \label{eq:Yukawa_CH}
\end{align} 
where $v =\sqrt{v_1^2+v_2^2}$ with $v_i$ being the VEV of neutral component of Higgs doublet $H_i$, the reduced Yukawa couplings $(\xi^{f,\ell}_\phi)_{ij}$ are given in \cite{Benbrik:2021wyl} in terms of the free parameters $\chi_{ij}^{f,\ell}$ and the mixing angle $\alpha$ of the two $\mathcal{CP}$-even scalar bosons and of $\tan\beta$.

\section{Results and discussion} 
In this contribution, as mentioned above we focus on both scenarios standard ($h\equiv h^{SM}$) and inverted hierarchy ($H\equiv h^{SM}$). We perform a random scan over the 2HDM parameter space using the public code \texttt{2HDMC-1.8.0} \cite{2hdmc},  we then require that each parameter point complies with the following theoretical and experimental constraints:
\begin{itemize}
\item \textbf{Theoretical constraints}: Unitarity \cite{uni1,uni2,uni3}, Perturbativity \cite{Branco:2011iw}, and Vacuum stability \cite{Barroso:2013awa,sta}.  
\item \textbf{Collider constraints}: Agreement with electroweak precision observables (EWPOs) \cite{Baak:2014ora} through the oblique parameters ($S$, $T$, $U$) \cite{Grimus:2007if,oblique2,Haller:2018nnx}, agreement  with current collider measurements of the Higgs signal strength as well as limits obtained from various searches of additional Higgs bosons at the LEP, Tevatron and LHC, in which we make use of  \texttt{HiggsBouns-5.9.0} \cite{HB} and  \texttt{HiggsSignal-2.6.0} \cite{HS}. and finally agreement with the current limits from B physics observables by using the public code \texttt{SuperIso-v4.1} \cite{superIso}.
\end{itemize}
\subsection{Scenario-I: Standard hierarchy}
\begin{figure}[H]
	\centering
	\includegraphics[height=10.5cm,width=15cm]{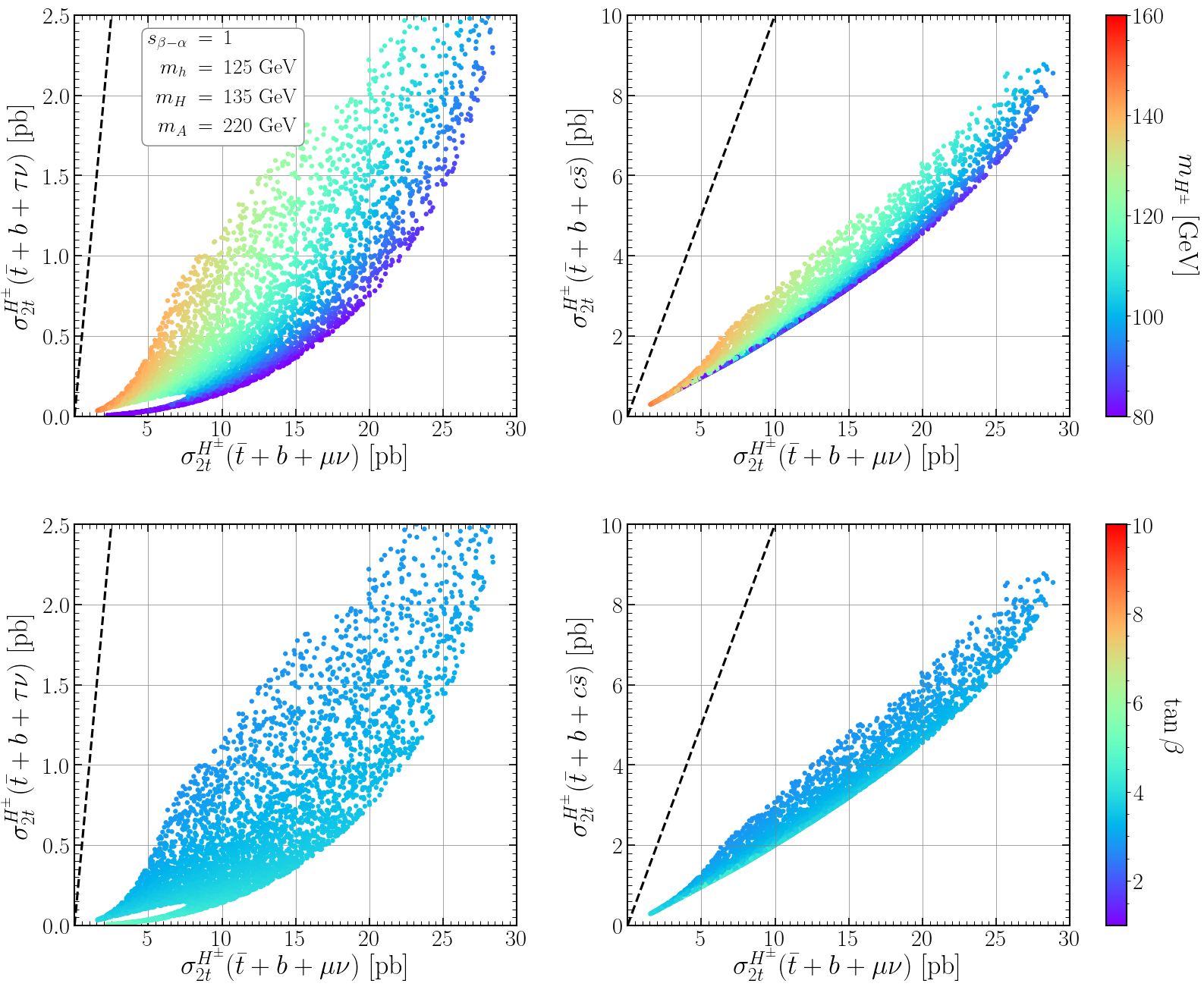}
	\caption{${\sigma^{H^{\pm}}_{2t}(\bar{t}+b+\mu\nu)}$ v.s ${\sigma^{H^{\pm}}_{2t}(\bar{t}+b+XY)}$ with XY$\equiv \tau\nu$ (left) and XY $\equiv c\bar{s}$ (right). The color bar shows the mass of the charged Higgs boson (up) and $\tan\beta$ (down).}\label{fig1}
\end{figure}
\subsection{Scenario-II: Inverted hierarchy}
\begin{figure}[H]
	\centering
	\includegraphics[height=10.5cm,width=15cm]{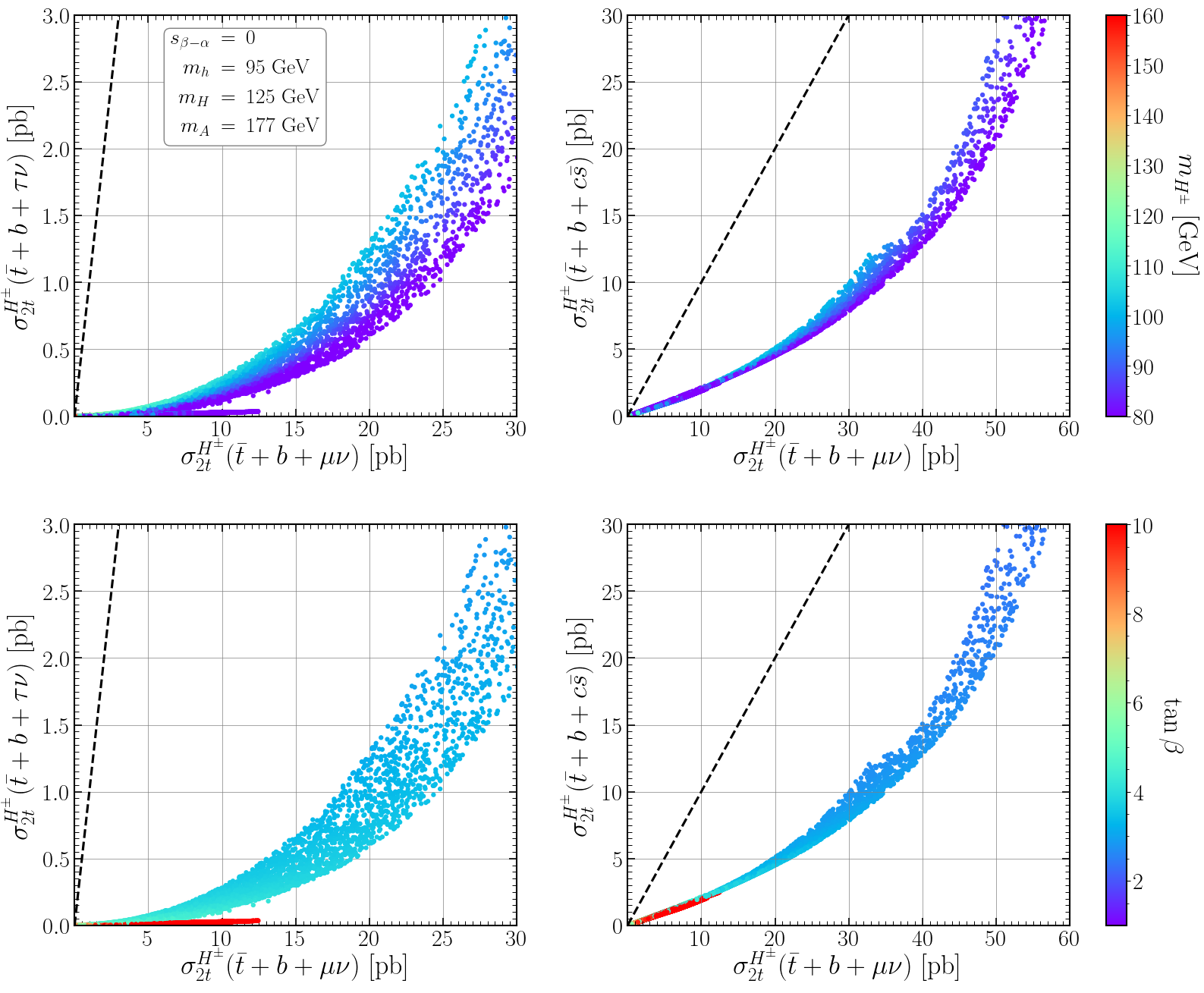}
	\caption{${\sigma^{H^{\pm}}_{2t}(\bar{t}+b+\mu\nu)}$ v.s ${\sigma^{H^{\pm}}_{2t}(\bar{t}+b+XY)}$ with XY$\equiv \tau\nu$ (left) and XY $\equiv c\bar{s}$ (right). The color bar shows the mass of the charged Higgs boson (up) and $\tan\beta$ (down).}\label{fig2}
\end{figure}
In Figs. \ref{fig1} and \ref{fig2}, we present the correlation between production and decay of light charged Higgs into $\mu\nu$ and $\tau\nu$ (left) and $c\bar{s}$ (right) mapping on charged Higgs mass $m_{H^\pm}$ (upper panels) and $\tan\beta$ (lower panels), for the parameter points that survive all constraints. It is clearly visible that, the signal cross section ${\sigma^{H^{\pm}}_{2t}(\bar{t}+b+\mu\nu)}$ is about one order of magnitude larger  over $\tau\nu$ and $c\bar{s}$ modes for roughly $m_{H^\pm}$ in 80 GeV $\sim$ 120 GeV range and $\tan\beta$ in 2 $\sim$ 5 range, This large production rate of $\mu\nu$ is due to the higher branching ratios of charged Higgs decay and its production rate from $pp \to t\bar{t}$ process. It is therefore clear that ${\sigma^{H^{\pm}}_{2t}(\bar{t}+b+\mu\nu)}$ can provide a potential alternative discovery channel for light charged Higgs boson at the LHC in the 2HDM Type-III.
\section{Conclusion}
In this contribution, we have investigated the production of light charged  Higgs bosons
via $pp \to tbH^\pm$ at the LHC with $\sqrt{s}$ = 13 TeV in both scenarios where $h$  and $H$ can identified to be the SM-like state of mass 125 GeV, Under the up-to-date theoretical and experimental constraints, we have mainly focused on the $\mu\nu$ decay channel of $H^\pm$, which is dominant over the other fermionic decay modes $\tau\nu$ and $c\bar{s}$. We have illustrated that such production
and decay channels would lead to significant cross section for $\mu\nu$ final states, which might be alternative signatures to target the light charged Higgs boson at the LHC.
\section*{Acknowledgments}
This work is supported by the Moroccan Ministry of Higher Education and Scientific Research MESRSFC and CNRST Project PPR/2015/6.


\begin{thebibliography}{99}

	\bibitem{Aad:2012tfa}
	\textbf{ATLAS} collaboration, G.~Aad \underline{et al.}, ``Observation of a new particle in the search for the Standard Model Higgs boson with the ATLAS detector at the LHC'', \href{https://www.sciencedirect.com/science/article/pii/S037026931200857X}{\underline{Phys. Lett.} \textbf{B716} (2012) 1-29}, \href{https://arxiv.org/pdf/1207.7214.pdf}{\texttt{arXiv:1207.7214 [hep-ex]}}.\vspace{0.15cm}	
	
	\bibitem{Chatrchyan:2012ufa}
	\textbf{CMS} collaboration, S.~Chatrchyan \underline{et al.}, ``Observation of a New Boson at a Mass of 125 GeV with the CMS Experiment at the LHC'', \href{https://www.sciencedirect.com/science/article/pii/S0370269312008581}{\underline{Phys. Lett.} \textbf{B716} (2012) 30-61}, \href{https://arxiv.org/pdf/1207.7235.pdf}{\texttt{arXiv:1207.7235 [hep-ex]}}.\vspace{0.15cm}
	
	
	\bibitem{Cheng:1987rs} 
	T.~P.~Cheng and M.~Sher, ``Mass Matrix Ansatz and Flavor Nonconservation in Models with Multiple Higgs Doublets'', \href{https://journals.aps.org/prd/abstract/10.1103/PhysRevD.35.3484}{\underline{Phys. Rev.} \textbf{D35} (1987) 3484}.\vspace{0.15cm}
	
	\bibitem{Benbrik:2021wyl}
	R.~Benbrik, M.~Boukidi, B.~Manaut, M.~Ouchemhou, S.~Semlali and S.~Taj, ``New charged Higgs boson discovery channel at the LHC'', \href{https://arxiv.org/pdf/2112.07502.pdf}{\texttt{arXiv:2112.07502 [hep-ph]}}.\vspace{0.15cm}
	
	\bibitem{Benbrik:2022azi}
	R.~Benbrik, M.~Boukidi, S.~Moretti and S.~Semlali,
	``Explaining the 96 GeV Di-photon Anomaly in a Generic 2HDM Type-III'', \href{https://arxiv.org/pdf/2204.07470.pdf}{\texttt{arXiv:2204.07470 [hep-ph]}}.\vspace{0.15cm}

	
	
	

   
   \bibitem{2hdmc}
   D.~Eriksson, J.~Rathsman and O.~Stal, ``2HDMC: Two-Higgs-Doublet Model Calculator Physics and Manual,''
   \href{https://www.sciencedirect.com/science/article/abs/pii/S0010465509003014?via/3Dihub}{\underline{Comput. Phys. Commun.} \textbf{181} (2010) 189-205}, \href{https://arxiv.org/pdf/0902.0851.pdf}{\texttt{arXiv:0902.0851 [hep-ph]}}.\vspace{0.15cm}
	
	\bibitem{uni1}
	S.~Kanemura, T.~Kubota and E.~Takasugi, ``Lee-Quigg-Thacker bounds for Higgs boson masses in a two doublet model'', \href{https://www.sciencedirect.com/science/article/abs/pii/0370269393912052?via/3Dihub}{\underline{Phys. Lett.} \textbf{B313} (1993) 155-160}, \href{https://arxiv.org/pdf/hep-ph/9303263.pdf}{\texttt{arXiv:hep-ph/9303263 [hep-ph]}}.\vspace{0.15cm}
	
	\bibitem{uni2}
	A.~G.~Akeroyd, A.~Arhrib and E.~M.~Naimi, ``Note on tree level unitarity in the general two Higgs doublet model'',  \href{https://www.sciencedirect.com/science/article/abs/pii/S037026930000962X?via/3Dihub}{\underline{Phys. Lett.} \textbf{B490} (2000) 119-124}, \href{https://arxiv.org/pdf/hep-ph/0006035.pdf}{\texttt{arXiv:hep-ph/0006035 [hep-ph]}}.\vspace{0.15cm}
	
	
	\bibitem{uni3}
	A.~Arhrib, ``Unitarity constraints on scalar parameters of the standard and two Higgs doublets model'', \href{https://arxiv.org/pdf/hep-ph/0012353.pdf}{\texttt{arXiv:hep-ph/0012353 [hep-ph]}}.\vspace{0.15cm}
	
	\bibitem{Branco:2011iw}
	G.~C.~Branco, P.~M.~Ferreira, L.~Lavoura, M.~N.~Rebelo, M.~Sher and J.~P.~Silva, ``Theory and phenomenology of two-Higgs-doublet models'', \href{https://www.sciencedirect.com/science/article/abs/pii/S0370157312000695?via/3Dihub}{\underline{Phys. Rept.} \textbf{516} (2012) 1-102}, \href{https://arxiv.org/pdf/1106.0034.pdf}{\texttt{arXiv:1106.0034 [hep-ph]}}.\vspace{0.15cm}
	
	\bibitem{Barroso:2013awa}
	A.~Barroso, P.~M.~Ferreira, I.~P.~Ivanov and R.~Santos, ``Metastability bounds on the two Higgs doublet model'', \href{https://doi.org/10.1007/JHEP06(2013)045}{\underline{JHEP} \textbf{06} (2013), 045}, \href{https://arxiv.org/pdf/1303.5098.pdf}{\texttt{arXiv:1303.5098 [hep-ph]}}.\vspace{0.15cm}
	
	
	\bibitem{sta}
	N.~G.~Deshpande and E.~Ma, ``Pattern of Symmetry Breaking with Two Higgs Doublets'', \href{https://journals.aps.org/prd/abstract/10.1103/PhysRevD.18.2574}{\underline{Phys. Rev.} \textbf{D18} (1978) 2574}.\vspace{0.15cm}
	
	
	
	
	
	
	\bibitem{Baak:2014ora}
	M.~Baak \textit{et al.} [Gfitter Group], ``The global electroweak fit at NNLO and prospects for the LHC and ILC'', \href{https://link.springer.com/article/10.1140/epjc/s10052-014-3046-5}{\underline{Eur. Phys. J.}  \textbf{C74} (2014), 3046}, \href{https://arxiv.org/pdf/1407.3792.pdf}{\texttt{arXiv:1407.3792 [hep-ph]}}.\vspace{0.15cm}
	
	
	\bibitem{Grimus:2007if}
	W.~Grimus, L.~Lavoura, O.~M.~Ogreid and P.~Osland, ``A Precision constraint on multi-Higgs-doublet models'', \href{https://iopscience.iop.org/article/10.1088/0954-3899/35/7/075001}{\underline{J. Phys.}  \textbf{G35} (2008), 075001}, \href{https://arxiv.org/pdf/0711.4022.pdf}{\texttt{arXiv:0711.4022 [hep-ph]}}.\vspace{0.15cm}
	
	
	
	\bibitem{oblique2}
	W.~Grimus, L.~Lavoura, O.~M.~Ogreid and P.~Osland, ``The Oblique parameters in multi-Higgs-doublet models'', \href{https://www.sciencedirect.com/science/article/abs/pii/S0550321308002289?via/3Dihub}{\underline{Nucl. Phys.}  \textbf{B801} (2008) 81-96}, \href{https://arxiv.org/pdf/0802.4353.pdf}{\texttt{arXiv:0802.4353 [hep-ph]}}.\vspace{0.15cm}
	
	
	\bibitem{Haller:2018nnx}
	J.~Haller, A.~Hoecker, R.~Kogler, K.~M\"onig, T.~Peiffer and J.~Stelzer, ``Update of the global electroweak fit and constraints on two-Higgs-doublet models'', \href{https://doi.org/10.1140/epjc/s10052-018-6131-3}{\underline{Eur. Phys. J.} \textbf{C78} (2018) no.8, 675}, \href{https://arxiv.org/pdf/1803.01853.pdf}{\texttt{arXiv:1803.01853 [hep-ph]}}.\vspace{0.15cm}
	
	\bibitem{HB}
	P.~Bechtle, D.~Dercks, S.~Heinemeyer, T.~Klingl, T.~Stefaniak, G.~Weiglein and J.~Wittbrodt, ``HiggsBounds-5: Testing Higgs Sectors in the LHC 13 TeV Era'', \href{https://link.springer.com/article/10.1140/2Fepjc/2Fs10052-020-08557-9}{\underline{Eur. Phys. J.}  \textbf{C80} (2020) 1211}, \href{https://arxiv.org/pdf/2006.06007.pdf}{\texttt{arXiv:2006.06007 [hep-ph]}}.\vspace{0.15cm}
	
	\bibitem{HS}
	P.~Bechtle, S.~Heinemeyer, T.~Klingl, T.~Stefaniak, G.~Weiglein and J.~Wittbrodt, ``HiggsSignals-2: Probing new physics with precision Higgs measurements in the LHC 13 TeV era'', \href{https://link.springer.com/article/10.1140/2Fepjc/2Fs10052-021-08942-y}{\underline{Eur. Phys. J.}  \textbf{C81} (2021) 145},  \href{https://arxiv.org/pdf/2012.09197.pdf}{\texttt{arXiv:2012.09197 [hep-ph]}}.\vspace{0.15cm}
	
	\bibitem{superIso}
	F.~Mahmoudi, ``SuperIso v2.3: A Program for calculating flavor physics observables in Supersymmetry'', \href{https://www.sciencedirect.com/science/article/abs/pii/S0010465509000721?via/3Dihub}{\underline{Comput. Phys. Commun.} \textbf{180} (2009) 1579-1613}, \href{https://arxiv.org/pdf/0808.3144.pdf}{\texttt{arXiv:0808.3144 [hep-ph]}}.\vspace{0.15cm}
\end{thebibliography}
\end{document}